\documentclass[nofootinbib,aps,preprint,superscriptaddress]{revtex4}
\usepackage{amsmath,amssymb}
\usepackage{citesort}
\usepackage{epsfig}

\newcommand{\beqa}{\begin{eqnarray}}
\newcommand{\eeqa}{\end{eqnarray}}
\newcommand{\beq}{\begin{equation}}
\newcommand{\eeq}{\end{equation}}
\newcommand{\bal}{\begin{align}}
\newcommand{\eal}{\end{align}}

\def\gsim{\ \rlap{\raise 3pt \hbox{$>$}}{\lower 3pt \hbox{$\sim$}}\ }
\def\lsim{\ \rlap{\raise 3pt \hbox{$<$}}{\lower 3pt \hbox{$\sim$}}\ }

\def\s{\sqrt{2}}

\begin{document}

\preprint{\vbox{
\hbox{}
\hbox{TECHNION-PH-2005-3}
\hbox{hep-ph/0502139}
\hbox{April 2005}
}}

\vspace*{48pt}
\title{Isospin-breaking effects on $\alpha$ 
extracted in $B\to\pi\pi,~\rho\rho,~\rho\pi$}

\def\addtech{Department of Physics,
Technion--Israel Institute of Technology,\\
Technion City, 32000 Haifa, Israel\vspace*{6pt}}

\author{M. Gronau}\affiliation{\addtech}
\author{J. Zupan}
\affiliation{Department of Physics, Carnegie Mellon University,\\
Pittsburgh, PA 15213\vspace*{6pt}}
\affiliation{J.~Stefan Institute, Jamova 39, P.O. Box 3000,1001
Ljubljana, Slovenia\vspace*{18pt}}

\begin{abstract} \vspace*{18pt}
Isospin-breaking in $B\to\pi\pi$ caused by $\pi^0-\eta-\eta'$ mixing is studied in 
a model-independent way using flavor SU(3).  Measured branching ratios for 
$B^+\to\pi^+\pi^0,~B^+\to~\pi^+\eta^{(')}$ and $B^0\to\pi^0\eta^{(')}$ imply an 
uncertainty in $\alpha$ smaller than $1.4^\circ$.
We find a negligible effect of $\pi^0-\eta-\eta'$ mixing on $\alpha$ in $B\to\rho\pi$. 
Characterizing the effect of $\rho^0-\omega$ mixing 
in $B\to\rho\rho$ and in $B\to\rho\pi$ by the two-pion invariant mass dependence, 
we point out a way of constraining this effect experimentally or eliminating it altogether. 
We show that a model-independent shift in $\alpha$ caused by electroweak penguin 
amplitudes in $B\to\pi\pi$ and $B\to\rho\rho$, $\Delta\alpha_{\rm EWP}=(1.5\pm 
0.3)^\circ$, may be slightly different in $B\to \rho\pi$. Other sources of isospin-breaking 
in these processes are briefly discussed.

\end{abstract}

\maketitle

\section{Introductiion}

Isospin symmetry provides triangle relations for $B\to\pi\pi$ and 
$\bar B \to\pi\pi$, which are governed by $I=0$ and $I=2$ amplitudes,
\beqa\label{triangle}
A_{+-} + \s A_{00} - \s A_{+0} & =  & 0~,~~~~~A_{ij} \equiv A(B^0 \to \pi^i \pi^j)~,\\
\bar A_{+-} + \s \bar A_{00} - \s \bar A_{-0} & =  & 0~,~~~~~\bar A_{ij} \equiv 
A(\bar B^0 \to \pi^i \pi^j)~.
\label{trianglebar}
\eeqa
These relations enable an extraction of the phase $\alpha\equiv \phi_2\equiv {\rm Arg}(-V^*_{tb}
V_{td}/V^*_{ub}V_{ud})$ from time-dependent CP 
asymmetries, $S_{\pi\pi}$ and $C_{\pi\pi}$,  in $B^0(t)\to \pi^+\pi^-$~\cite{GL}.
The asymmetries determine $\sin 2\alpha_{\rm eff}$,
\beq
\sin(2\alpha_{\rm eff}) = \frac{S_{\pi\pi}}{\sqrt{1 - C^2_{\pi\pi}}}~.
\eeq
The shift $\Delta\alpha\equiv \alpha_{\rm eff} - \alpha$ caused by the penguin 
amplitude is given by 
\beq\label{theta}
\Delta\alpha \equiv  \frac{1}{2}{\rm Arg}\left (e^{2i\gamma}
\bar A_{+-}A^*_{+-}\right )~.
\eeq

We define a measurable phase $\Delta\alpha_0$, given in terms of
angles in the $B$ and $\bar B$ triangles, $\phi \equiv  {\rm Arg}\left 
(A_{+-}A^*_{+0}\right )$, $\bar\phi \equiv {\rm Arg}\left (\bar A_{+-}
\bar A^*_{-0}\right)$: 
\beq\label{alpha0}
\Delta\alpha_0 \equiv \frac{1}{2}(\bar\phi - \phi)
= \frac{1}{2}\left [{\rm Arg}\left (e^{2i\gamma}\bar A_{+-}A^*_{+-}\right )
- {\rm Arg}\left (e^{2i\gamma}\bar A_{-0}A^*_{+0}\right )\right ]~.
\eeq
Neglecting very small electroweak penguin amplitudes which will be 
discussed below, a phase relation holds between the two $\Delta I=3/2$ 
tree amplitudes, 
\beq\label{A+0phase}
A_{+0} = e^{2i\gamma}\bar A_{-0}~.
\eeq
This implies $\Delta\alpha = \Delta\alpha_0$, fixing the relative orientation 
of the $B$ triangle and the $\bar B$ triangle (rotated by $2\gamma$) 
such that these sides overlap.  In this configuration $\Delta\alpha$ is half the angle 
between sides corresponding to $B^0\to\pi^+\pi^-$ and $\bar B^0\to\pi^+\pi^-$.
This determines $\Delta\alpha$ up to a fourfold ambiguity (including a sign 
ambiguity) related to the four possible relative orientations of the two triangles.
In case that $|A(B^0\to\pi^0\pi^0)|$ and $|A(\bar B^0\to\pi^0\pi^0)|$ are not
separately measured, while the charge-averaged neutral pion rate is measured, 
one may obtain upper bounds on $|\Delta\alpha|$~\cite{bounds}.

The same method applies to polarization states in $B^0(t)\to \rho^+\rho^-$
which are CP-eigenstates, in particular to an even-CP longitudinally polarized state 
which is found to dominate this process~~\cite{polBa}.
A variant of  isospin symmetry can also be used to learn $\alpha$  in 
$B^0(t) \to \rho\pi$~\cite{rhopi,QS}. Application of these 
methods to recent measurements by BaBar~\cite{polBa,BaBar} and 
Belle~\cite{Belle}, where $B\to \rho\rho$ played a dominant role, provides 
the currently most accurate direct determination of 
$\alpha$~\cite{CKMfitter,SU3}, $\alpha = (100^{+9}_{-10})^\circ$. 
This precision can be improved~\cite{GLW} by resolving the sign ambiguity 
in $\Delta\alpha$ under reasonably mild and testable assumptions about 
strong phase differences between tree and penguin amplitudes in these 
processes. 
The error in $\alpha$ is expected to be reduced further by improving the 
measurement of the direct CP asymmetry in $B^0\to\pi^0\pi^0$~\cite{C00}, 
or by using a prediction of a SCET analysis~\cite{BPRS} that the phase 
difference between tree and color-suppressed amplitudes in 
$B^+\to\pi^+\pi^0$ is small~\cite{BRS}. 

At this level of precision one is required to consider small 
electroweak penguin (EWP) amplitudes and corrections from isospin 
breaking caused by the $u$ and $d$ charge and mass differences. These 
corrections  modify the geometry of the $B$ and $\bar B$ amplitude 
triangles. A model-independent study of electroweak penguin contributions 
in $B\to\pi\pi$ was performed in~\cite{EWP}. Instead of overlapping with 
each other, the two sides of the two triangles, $A_{+0}$ and  
$e^{2i\gamma}\bar A_{-0}$, were shown to form a calculable relative angle.
Neglecting EWP operators with small Wilson coefficients ($c_7$ and $c_8$), 
isospin symmetry relates dominant $\Delta I =3/2$ EWP operators to $\Delta I=3/2$
current-current operators in the effective Hamiltonian, implying
\beqa\label{DeltathetaEWP}
(\Delta\alpha - \Delta\alpha_0)_{\rm EWP} & \equiv & \frac{1}{2}{\rm Arg}
(e^{2i\gamma}\bar A_{-0}A^*_{+0})
= -\frac{3}{2}\left (\frac{c_9 + c_{10}}{c_1 + c_2}\right )\frac{|V_{tb}V_{td}|}
{|V_{ub}V_{ud}|}\sin\alpha
\nonumber\\
& = & -\frac{3}{2}\left (\frac{c_9 + c_{10}}{c_1 + c_2}\right )\frac{\sin(\beta 
+ \alpha)\sin\alpha}{\sin\beta} = +0.013 \frac{\sin(\beta + \alpha)\sin\alpha}
{\sin\beta}~.
\eeqa
The measured values of $\beta$ and $\alpha$~\cite{CKMfitter}, 
$\beta = (23.3 \pm 1.6)^\circ,~\alpha = (100^{+9}_{-10})^\circ$, lead to a small 
calculable value with a negligible error, $(\Delta\alpha - \Delta\alpha_0)_{\rm EWP} = 
(1.5\pm 0.3)^\circ$. This shift must be included in the determination of $\alpha$ 
by using $\alpha = \alpha_{\rm eff} - \Delta\alpha_0 - (\Delta\alpha - 
\Delta\alpha_0)_{\rm EWP}$. 

Isospin-breaking due to nonzero $u$ and $d$ quark mass and charge differences 
has several effects on the analysis of $B\to\pi\pi$, $B\to\rho\rho$ and $B\to\rho\pi$. 
An important effect in $B\to\pi\pi$, caused by $\pi^0- \eta-\eta'$ mixing, was studied 
several years ago by Gardner~\cite{Gardner} using 
generalized factorization~\cite{AKL}. She concluded
that the resulting error on the extracted value of $\alpha$ in the above range is about
 $5^\circ$ including EWP contributions. 
The uncertainty may be even larger due to the approximation involved in 
this estimate. This would limit severely the future accuracy of 
determining $\alpha$ in $B\to\pi\pi$. 
Gardner and Meissner~\cite{GM} discussed briefly 
the appearance of a small $\Delta I =5/2$ amplitude in $B\to\pi\pi$ which violates 
the isospin triangle relation. 
They mentioned isospin violation in $B\to\rho\pi$ caused by $\pi^0- \eta-\eta'$ mixing 
and by $\rho-\omega$ mixing, pointing out that the presence of an additional 
$\Delta I=5/2$ amplitude in $B\to\rho\pi$,
involving  the same weak phase as the tree amplitude,  does not affect the isospin analysis.  
Refs.~\cite{ET,GOT,LGT} studied 
direct CP violation in $B^{+,0}\to (\pi^+\pi^-)_{\rho,\omega}\pi^{+,0}$ and in
$B^+\to (\pi^+\pi^-)_{\rho,\omega}\rho^+$ caused by $\rho-\omega$ mixing. These CP 
asymmetries affect the analyses of isospin related processes. Ref.~\cite{CKMfitter} 
studied numerically the uncertainty in determining $\alpha$ in $B\to\rho\rho$, assuming 
that isospin violating corrections in tree and penguin amplitudes in 
$\s A(B^+\to\rho^+\rho^0)$ are at a level of 4$\%$ relative to tree and penguin 
amplitudes in $B^0\to\rho^+\rho^-$.   

The purpose of this work is to analyze in a model-independent manner
isospin-breaking effects, in particular the effects of $\pi^0- \eta-\eta'$ mixing 
and $\rho-\omega$ mixing, on determining 
$\alpha$ in $B\to\pi\pi,~B\to\rho\rho$ and $B\to\rho\pi$.  In Section II we will 
apply flavor SU(3) to $B$ decays into two charmless pseudoscalars, 
relating isospin-breaking terms in $B\to\pi\pi$ to amplitudes of $B\to \pi\eta$ and 
$B\to \pi\eta'$. Using measured rates, we will show that the effect of $\pi^0-\eta-\eta'$ 
mixing on determining $\alpha$ in $B\to\pi\pi$ is considerably smaller than estimated 
by Gardner. 
Turning in Section III to discuss the effects of $\rho-\omega$ mixing on determining $\alpha$ in 
$B\to\rho\rho$, we will show how to include this effect experimentally, without having to
rely on a calculation of $\rho-\omega$ mixing parameters. Section IV studies $\pi^0-\eta-\eta'$  
mixing, $\rho-\omega$ mixing 
and the effect of electroweak penguin amplitudes in $B\to\rho\pi$.
In Section V we discuss briefly other sources for isospin-breaking,
while Section VI concludes. Appendix A presents 
experimental constraints on parameters describing $\rho-\omega$ mixing.

\section{Effect of $\pi^0-\eta-\eta'$ mixing in $B\to\pi\pi$}

The mixing of $\pi^0$ with $\eta$ and $\eta'$ introduces isospin-breaking  
in $B\to\pi\pi$ through an additional $I=1$ amplitude, while the isospin 
conserving terms obey the triangle 
relation (\ref{triangle}). We will use flavor SU(3) symmetry to estimate the 
isospin-breaking terms. SU(3) breaking corrections and smaller annihilation-like 
amplitudes which we neglect in these terms are higher order, and are expected to 
introduce an uncertainty at a level of 30$\%$. A convenient way of 
applying flavor SU(3) 
to charmless $B$ decays into two pseudoscalars is in terms of graphical 
representations describing flavor flow topologies~\cite{DZ,Chau,GHLR,DGR}. 
SU(3) amplitudes for two octets in the final state consists of a ``tree" amplitude 
($t$) a ``color-suppressed" amplitude ($c$) a ``penguin" amplitude ($p$).
Three annihilation-like amplitudes ($a,~e$ and $pa$) are expected to be much 
smaller~\cite{GHLR, BP} and will be neglected. The remaining three amplitudes 
contain EWP contributions~\cite{GHLREWP}, the overall effect of which can be 
taken into account as summarized in Eq.~(\ref{DeltathetaEWP}). This effect can 
be included as explained above, and will therefore be disregarded in this section.
For a singlet and an octet in the final state one has three SU(3) 
amplitudes~\cite{DZ}, of which a ``singlet penguin" amplitude ($s$) dominates, 
while two annihilation-like amplitudes will be neglected~\cite{DGR}.

We  use quark content for mesons as in \cite{GHLR,DGR}, but a somewhat 
different  phase convention:
\beqa\label{convention}
B^0 & = & -d\bar b,~~B^+ = -u\bar b,~~\pi_3=\frac{1}{\s}(u\bar u-d\bar d),~
\pi^+ = -u\bar d,~~\pi^- = d\bar u,\nonumber\\
\eta & = & \frac{1}{\sqrt{3}}(u\bar u+d\bar d-s\bar s),~~~~~
\eta'=\frac{1}{\sqrt{6}}(u\bar u+d\bar d+2s\bar s)~.
\eeqa
The $\eta$ and $\eta'$ correspond to octet-singlet mixtures,
\beqa
\eta  & = & \eta_8 \cos \theta - \eta_1 \sin \theta~,~~
\eta' = \eta_8 \sin \theta + \eta_1 \cos \theta~,\nonumber\\
\eta_8 & = & \frac{1}{\sqrt{6}}(u\bar u + d\bar d - 2s\bar s)~,~~
\eta_1 = \frac{1}{\sqrt{3}}(u\bar  u + d\bar d + s\bar s )~,
\eeqa
with an ``ideal" mixing angle $\theta=\theta_0 = \sin^{-1}(-1/3) = -19.5^\circ$. A slightly 
larger magnitude, $\theta = -22^\circ$,
was used in~\cite{Gardner}, while a slightly smaller magnitude, $\theta = 
(-15.7\pm 1.7)^\circ$, was obtained in a very recent phenomenological 
fit~\cite{EF}. While in the most part we use the value $\theta_0$, at the end of this section
we discuss briefly the effect of a variation in $\theta$.

Expressions for decay amplitudes in terms of graphical SU(3) contributions, for final 
states involving pairs of isotriplet pions, and for pairs involving an isotriplet pion 
and $\eta$ or 
$\eta'$,  are obtained in a straightforward manner~\cite{GHLR,DGR,piKeta}:
\beqa\label{+-3}
& & A_{+-} = t+p~,~~~~A_{33} = \frac{1}{\s}(c-p)~,~~~~A_{+3} = \frac{1}{\s}(t+c)~,
\nonumber\\
& & A_{3\eta} = \frac{1}{\sqrt{6}}(2p+s)~,~~~~~~~~~~~
~~~~~~~~~~~~~~~~~~A_{3\eta'} = \frac{1}{\sqrt{3}}(p+2s)~,
\nonumber\\
& & A_{+\eta} = \frac{1}{\sqrt{3}}(t+c+2p+s)~,~~~~~~~~
~~~~~~~~A_{+\eta'} = \frac{1}{\sqrt{6}}(t+c+2p+4s)~.
\eeqa
The first three amplitudes for pure isotriplet pions obey clearly the isospin triangle relation 
(\ref{triangle}). The purely $\Delta I =3/2$ amplitude $A_{+3}=(t+c)/\s$ has a weak 
phase $\gamma$.

The mixing of $\pi^0,~\eta$ and $\eta'$ introduces a small isospin singlet 
component into the dominantly isotriplet neutral pion state,
\beq\label{mixing}
|\pi^0\rangle=|\pi_3\rangle +\epsilon |\eta\rangle +\epsilon'|\eta'\rangle~.
\eeq
Values $\epsilon=0.014,~\epsilon'=0.0077$ were used 
by Gardner~\cite{Gardner}, based on a calculation applying chiral perturbation 
theory~\cite{Leutwyler}. We will take ranges of values 
as obtained in a recent update~\cite{Kroll}, 
$\epsilon = 0.017 \pm 0.003,~\epsilon' = 0.004 \pm 0.001$. 
 
Neglecting terms quadratic in $\epsilon$ and $\epsilon'$, we find decay amplitudes for
the neutral pion state given by (\ref{mixing}):
\beqa
A_{+0} & = & A_{+3} + \epsilon A_{+\eta} + \epsilon' A_{+\eta'} \nonumber\\
& = & \frac{1}{\s}(t+c)(1+e_0) + \frac{1}{\sqrt{3}}\epsilon (2p+s) + \frac{\s}{\sqrt{3}}\epsilon'(p+2s)~,\\
A_{00} & = & A_{33} + \s \epsilon A_{3\eta} + \s \epsilon' A_{3\eta'}\label{00} \nonumber\\
& = & \frac{1}{\s}(c-p) + \frac{1}{\sqrt{3}}\epsilon (2p+s) + \frac{\s}{\sqrt{3}}\epsilon'(p+2s)~,
\eeqa
where
\beq\label{e_0}
e_0 = \sqrt{\frac{2}{3}}\epsilon + \sqrt{\frac{1}{3}}\epsilon' = 0.016\pm 0.003~.
\eeq
We note the factors $\s$ in the first line of Eq.~(\ref{00}).
This takes into account final states of identical particles in $A_{00}$ and $A_{33}$~\cite{identical}, 
compared to states which must be symmetrized in $A_{3\eta}$ and $A_{3\eta'}$. Squares of 
amplitudes give decay rates when common phase space factors are implied.

Using these expressions, we find two important consequences of this amplitude 
decomposition which includes $\pi^0-\eta-\eta'$ mixing:
\begin{enumerate}
\item The triangle relation (\ref{triangle}) is modified only slightly:
\beq\label{triangle2}
A_{+-} + \s A_{00} - \s A_{+0} (1-e_0) =  0~.
\eeq
\item The amplitude $A_{+0}$ can be written in terms of the pure $\Delta I=3/2$ 
amplitude, $A_{+3}$, carrying a weak phase $\gamma$, corrected by isospin-breaking 
terms involving $A_{0\eta}$ and $A_{0\eta'}$,
\beq\label{phaseA+0}
A_{+0} = A_{+3}(1+e_0) + \s\epsilon A_{0\eta} + \s\epsilon' A_{0\eta'}~.
\eeq
\end{enumerate}

Our first conclusion is therefore that the physical $B\to\pi\pi$ and $\bar B\to\pi\pi$ decay 
amplitudes still obey triangle relations. The isospin-breaking 
factor, $1-e_0$, multiplying the amplitude $A_{+0}$ in (\ref{triangle2}), can be absorbed in this 
measurement. Since $e_0$ is calculated to be between one and two percent, while the 
current error in $|A_{+0}|$ is about $5\%$ (see Eq.~(\ref{B0exp}) below), the factor 
$1-e_0$ starts to play a non-negligible role and must be included in the construction of the 
isospin triangles~\cite{Delta-phi}. The remaining error
from the theoretical uncertainty in $e_0$ given in (\ref{e_0}) is only a fraction of a percent,
causing a negligible error in determining $\Delta\alpha_0$ from the angles in 
the two isospin triangles.


The second result, Eq.~(\ref{phaseA+0}), implies that $A_{+0}$ and its 
charge-conjugate no longer obey the exact phase relation (\ref{A+0phase}). That is, since 
the weak phases of the small isospin-breaking terms in (\ref{phaseA+0}) differ from 
$\gamma$, the triangle (\ref{triangle2}) and the corresponding triangle for 
$\bar B$ amplitudes rotated by an angle $2\gamma$ do not share 
exactly a common base, $A_{+0} \ne e^{2i\gamma}\bar A_{-0}$. Denoting 
\beq
\psi_{\eta^{(')}} \equiv {\rm Arg}\left [\frac{A_{0\eta^{(')}}}{A_{+0}}\right ]~,~~~~~
\bar \psi_{\eta^{(')}} \equiv {\rm Arg}\left [\frac{\bar A_{0\eta^{(')}}}{\bar A_{-0}}\right ]~,
\eeq
this introduces a change, $\Delta\alpha - \Delta\alpha_0$, given to first order  
in $\epsilon$ and $\epsilon'$ by
\beqa\label{theta-function}
(\Delta\alpha & - & \Delta\alpha_0)_{\pi-\eta-\eta'} \equiv \frac{1}{2}
{\rm Arg}(e^{2i\gamma}\bar A_{-0}A^*_{+0})
\\
& = & \frac{1}{\s|A_{+0}|} \Big [\epsilon  
\Big(|\bar A_{0\eta}| \sin\bar\psi_{\eta} - |A_{0\eta}| \sin\psi_{\eta}\Big)
+\epsilon' \Big(|\bar A_{0\eta'}| \sin\bar\psi_{\eta'}-|A_{0\eta'}| \sin\psi_{\eta'}\Big)\Big ]~.
\nonumber
\eeqa

Given that the phases $\psi_{\eta^{(')}} $ and $\bar \psi_{\eta^{(')}} $ are unknown, 
an immediate upper bound on $|\Delta\alpha - \Delta\alpha_0|$ may be obtained by 
taking $\bar \psi_{\eta^{(')}}= -\psi_{\eta^{(')}}= \pi/2$: 
\beqa\label{bound}
|(\Delta\alpha - \Delta\alpha_0)_{\pi-\eta-\eta'}| & \le & \epsilon\left (\frac{|A_{0\eta}|+|\bar A_{0\eta}|}{\s|A_{+0}|} \right )
+ \epsilon' \left (\frac{|A_{0\eta'}|+|\bar A_{0\eta'}|}{\s|A_{+0}|}\right ) \nonumber\\
& \le & \sqrt{2\frac{\tau_+}{\tau_0}}\left (\epsilon\sqrt{\frac{{\cal B}_{0\eta}}{{\cal B}_{+0}}}
+ \epsilon' \sqrt{\frac{{\cal B}_{0\eta'}}{{\cal B}_{+0}}} \right )~.
\eeqa
Here ${\cal B}_{ij}\equiv (|A_{ij}|^2 + |\bar A_{ij}|^2)\tau_B/2$ denote charge-averaged
 branching ratios for corresponding decays, and $\tau_+/\tau_0$ is the lifetime ratio of 
 $B^+$ and $B^0$. We neglect tiny corrections (at a level of a percent) in phase space 
 factors. 

Using world averaged values~\cite{HFAG},
\beqa\label{B0exp}
\frac{\tau_+}{\tau_0}   & = & 1.081 \pm 0.015~,~~~~~~~~~~~~~~~~~~~~~~~
{\cal B}_{+0} = (5.5 \pm 0.6)\times 10^{-6}~\mbox{\cite{Be+0,Ba+0}}~,
\nonumber\\
{\cal B}_{0\eta} & < & 2.5\times 10^{-6}~(90\%~{\rm CL})~\mbox{\cite{Ba0eta,Beeta}}~,~~~
{\cal B}_{0\eta'}  <  3.7\times 10^{-6}~(90\%~{\rm CL})~\mbox{\cite{Ba0eta}}~,
\eeqa
we find at $90\%$ CL
\beq\label{bound'}
|(\Delta\alpha - \Delta\alpha_0)_{\pi-\eta-\eta'}|  < 1.05\epsilon + 1.28\epsilon' = 1.6^\circ~.
\eeq

The phases $\psi_{\eta} $ and $\psi_{\eta'}$ may actually be measured within 
discrete ambiguities through two triangle relations implied by Eqs.~(\ref{+-3}),
valid to zeroth order in $\epsilon$ and $\epsilon'$,
 \beqa\label{triangles-eta}
A_{+\eta} & = & \frac{\sqrt{2}}{\sqrt{3}}A_{+0}+\sqrt{2} A_{0\eta}~, \nonumber\\
A_{+\eta'} &= & \frac{1}{\sqrt{3}}A_{+0}+\sqrt{2} A_{0\eta'}~.
\eeqa
Measuring the magnitudes of the three amplitudes in each 
of the two triangles determines  
$\cos\psi_{\eta}$ and $\cos\psi_{\eta'}$. Similar relations for charge-conjugate 
amplitudes determine $\bar \psi_{\eta}$ 
and  $\bar \psi_{\eta'}$. To determine separately $|A_{0\eta^{(')}}|$ and 
$|\bar A_{0\eta^{(')}}|$ would require measuring also CP asymmetries in these channels.
In the absence of these asymmetry measurements, one may use charge-averaged rates alone 
to improve the upper bound (\ref{bound}). Maximizing $(\Delta\alpha-\Delta\alpha_0)_{\pi-\eta-\eta'}$ 
in (\ref{theta-function}) by 
varying $\psi_{\eta^{(')}}$ and $\bar\psi_{\eta^{(')}}$, while keeping  ${\cal B}_{+\eta^{(')}}$ and the
upper bounds on ${\cal B}_{0\eta^{(')}}$ fixed, we find that a maximum is obtained for 
$|\bar A_{0\eta^{(')}}| = |A_{0\eta^{(')}}|$: 
\beq
|(\Delta\alpha - \Delta\alpha_0)_{\pi-\eta-\eta'}|  \le  \sqrt{2\frac{\tau_+}{\tau_0}}
\left (\epsilon\sqrt{\frac{{\cal B}_{0\eta}}{{\cal B}_{+0}}(1 - r_{\eta})}
+ \epsilon' \sqrt{\frac{{\cal B}_{0\eta'}}{{\cal B}_{+0}}(1 - r_{\eta'})} \right )~,
\eeq
where
\beq
r_{\eta} = \frac{3}{16}\frac{\left [ \sqrt{\frac{\tau_0}{\tau_+}}({\cal B}_{+\eta} -
 \frac{2}{3}{\cal B}_{+0})
 - 2\sqrt{\frac{\tau_+}{\tau_0}}{\cal B}_{0\eta}\right ]^2}
{{\cal B}_{+0}{\cal B}_{0\eta}}~,~~~~
r_{\eta'} = \frac{3}{8}\frac{\left [ \sqrt{\frac{\tau_0}{\tau_+}}({\cal B}_{+\eta'} - 
\frac{1}{3}{\cal B}_{+0} )
- 2\sqrt{\frac{\tau_+}{\tau_0}}{\cal B}_{0\eta'}\right ]^2}
{{\cal B}_{+0}{\cal B}_{0\eta'}}~.
\eeq 

Using world averaged values~\cite{HFAG},
\beq\label{B+exp}
{\cal B}_{+\eta} = (4.8 \pm 0.6)\times 10^{-6}~\mbox{\cite{Ba+eta,Beeta}}~,~~~~~
{\cal B}_{+\eta'} = (4.2 \pm 1.1)\times 10^{-6}~\mbox{\cite{Ba+eta'}}~,
\eeq
and values in (\ref{B0exp}), we find at 90\% CL
\beq\label{bound'2}
|(\Delta\alpha - \Delta\alpha_0)_{\pi-\eta-\eta'}|  < 
1.4^\circ~.
\eeq
This is only a slight  improvement relative to (\ref{bound'}).

The  upper bounds (\ref{bound'}) and (\ref{bound'2})
involve an uncertainty of about 30$\%$ from SU(3) breaking and small annihilation 
amplitudes which we have neglected. The bounds are seen to be considerably lower 
than the estimate of the uncertainty, $\delta\alpha \sim 5^\circ$, obtained in~\cite{Gardner} 
using generalized 
factorization. These bounds may be tightened further by reducing errors in the relevant $B^+$ decay branching ratios, 
and in particular by improving the upper limits on ${\cal B}(B\to \pi^0\eta)$ 
and ${\cal B}(B \to \pi^0\eta')$. These experimental upper limits play also an important 
role in interpreting theoretically~\cite{GRZ} the measured deviation of the time-dependent 
CP asymmetry in $B^0\to\eta' K_S$ from $\sin 2\beta\sin\Delta mt$~\cite{eta'K}.
This makes the case for their further improvement even stronger.

Our analysis was based on the ``ideal" mixing angle, $\theta_0=\sin^{-1}(-1/3)$, which 
we used in (\ref{convention}). 
Defining a general mixing angle, $\theta = \theta_0 + \delta$, one may show that for
this case one must replace $e_0$ in (\ref{triangle2})  and (\ref{phaseA+0}) by $e$,
\beq
e = \sqrt{\frac{2}{3}}(\epsilon\cos\delta + \epsilon'\sin\delta) + 
\sqrt{\frac{1}{3}}(-\epsilon\sin\delta + \epsilon'\cos\delta)~.
\eeq
The small theoretical uncertainty in the value of $\theta$ ($|\delta| < 6^\circ$) implies a 
value for $e$ within the uncertainty  in $e_0$  given in (\ref{e_0}).
Furthermore, the terms $\epsilon$ and $\epsilon'$ in (\ref{phaseA+0}) are preserved 
by replacing $\theta_0$ by $\theta$, implying that the upper bounds (\ref{bound'}) and 
(\ref{bound'2}) are unaffected by varying $\theta$.
 
\section{Effects of $\rho-\omega$ mixing in $B\to \rho\rho$}

The processes $B\to \rho^i(\pi_1\pi_2)\rho^j(\pi_3\pi_4)$ are quasi two-body decays involving four 
pions in the  final state. To account for the $\rho$ width, 
the two $\rho$ mesons are defined by choosing suitable common ranges of invariant masses
for the two-pion pairs,
$s_{12}\equiv (p_1+p_2)^2$ and $s_{34}\equiv (p_3 + p_4)^2$. 
One uses the pion angular distributions in the $\rho$ rest frames to project 
longitudinally polarized states which were 
shown to dominate $B\to\rho\rho$~\cite{polBa}. Applying the isospin analysis to 
$B\to (\rho\rho)_{\rm long}$ proceeds identically to $B\to\pi\pi$~\cite{GL,bounds} in the limit of a vanishing $\rho$ width. (In principle, the method applies separately to each transversity state.)
The $\rho$ width has the effect that two $\rho$ mesons with different invariant masses,
$s_{12}\ne s_{34}$, cannot be considered identical. Therefore Bose symmetry does not 
exclude a final $I=1$ state~\cite{FLNQ}, for which the amplitude is antisymmetric under
$s_{12} \leftrightarrow s_{34}$. This amplitude does not interfere in the decay rate with the 
usual symmetric $I=0$ and $I=2$ amplitudes. The effect of the $I=1$ amplitude on the isospin 
analysis, of order $(\Gamma_\rho/m_\rho)^2 \simeq 0.04$, may be taken into account by 
including it in the fit. In principle,  the effect may be eliminated 
by decreasing the width of the $\rho$ band, however  this would also decrease the statistics.

In the following we will disregard this $I=1$ term, which contributes also in the isospin 
symmetry limit, studying isospin-breaking effects of the same order. 
To make our point, consider first the general invariant mass
dependence of decay amplitudes for the three distinct charged $\rho$ states,
\beqa\label{A+-0}
A_{+-}(s_{12},s_{34})  & \equiv & A(B^0\to(\pi^+\pi^0)_{12}(\pi^0\pi^-)_{34})
=A(B^0\to \rho^+\rho^-)f_c(s_{12})f_c(s_{34})~,
\nonumber\\
A_{+0}(s_{12},s_{34}) & \equiv & A(B^+\to(\pi^+\pi^0)_{12}(\pi^+\pi^-)_{34})
=A(B^+\to \rho^+\rho^0)f_c(s_{12})f_n(s_{34})~,
\nonumber\\
A_{00}(s_{12},s_{34}) & \equiv & A(B^0\to(\pi^+\pi^-)_{12}(\pi^+\pi^-)_{34})
=A(B^0\to \rho^0\rho^0)f_n(s_{12})f_n(s_{34})~,
\eeqa 
where $f_{c,n}$ are usually taken as Breit-Wigner factors.
If isospin symmetry were exact, then $f_n(s)=f_c(s)$, so that the two  ratios,
\beq\label{2ratios}
\frac{A_{+0}(s_{12},s_{34}) }{A_{+-}(s_{12},s_{34})}~~~~~~~{\rm and}~~~~~~~
\frac{A_{00}(s_{12},s_{34})}{A_{+-}(s_{12},s_{34})}~,
\eeq
would be independent of $s_{12}$ and $s_{34}$ in the quasi two-body approximation.
Any observed dependence of these ratios on the invariant masses would indicate either 
isospin-breaking, or dependence of $A(B\to \rho^i\rho^j)$ on $s_{12}$ and $s_{34}$. 
The latter possibility may be fitted experimentally by considering this dependence over 
the entire widths of the two $\rho$ mesons~\cite{FLNQ}. We will study isospin-breaking 
in a narrow range of invariant masses defined by the narrow $\omega$ resonance, for 
which $A(B\to \rho^i\rho^j)$ may be assumed to be constant.
Our purpose is to use the measured invariant mass dependence in (\ref{A+-0}) as a tool for 
extracting the isospin symmetric $B\to\rho\rho$ amplitudes which obey a triangle relation 
similar to (\ref{triangle}).

Let us now study $\rho-\omega$ mixing following a formalism developed in ~\cite{OPTW}.
The physical $\rho$ and $\omega$ fields are mixtures of an isovector field, $\rho_{I}$, and 
an isoscalar fields, $\omega_{I}$,
\beq\label{rho-mix}
\begin{split}
\rho^0&=\rho_{I}-\epsilon_1 \omega_{I}~,\\
\omega&=\omega_{I}+\epsilon_2 \rho_{I}~.
\end{split}
\eeq
The isospin-breaking parameters,  $\epsilon_{1,2}$, are of order of a few percent. A precise
knowledge of their magnitudes will not be needed (see appendix \ref{appA} for current  
experimental constraints), as they will be hidden in an isospin-breaking 
function to be introduced below. An expansion in $\epsilon_{1,2}$ will be carried out to first
order in these parameters.

Consider the transformation between the isospin basis and the physical basis for
the scalar parts of the vector meson propagators. 
The mixed propagator in the isospin basis, $D^I_{\rho\omega}\equiv \langle \rho_{I} \omega_{I}\rangle_0$, 
has poles at the $\rho$ and $\omega$ masses, 
and is conventionally written in the form
\beq\label{Drho-omega}
D_{\rho\omega}^I(s)=\Pi_{\rho\omega}(s) D_{\rho\rho}(s) D_{\omega\omega}(s)~.
\eeq
The physical basis is defined by requiring that 
$\Pi_{\rho\omega}$ does not have poles. 
The scalar parts of the physical propagators can be approximated near the poles by 
Breit-Wigner forms,
\beq\label{BW}
D_{VV}(s)=\frac{1}{s-m_V^2+i  m_V \Gamma_V}~,~~~~~V=\rho, \omega~.
\eeq 
The values of 
$\epsilon_{1,2}$ are chosen such that the mixed propagator in the physical basis, 
$D_{\rho\omega}\equiv \langle \rho \omega\rangle_0$ has  no poles,
\beq
D_{\rho\omega}^I  =  D_{\rho\omega}+\epsilon_1 D_{\omega\omega}-
\epsilon_2D_{\rho\rho}~.
\eeq
All three terms on the right-hand-side are of order $\epsilon_{1,2}$.
The equalities, $D^I_{VV}(s) = D_{VV}(s)$, ($V=\rho, \omega$), hold to first order in 
$\epsilon_{1,2}$. For instance, the second and third terms in the relation, 
\beq
D_{\rho\rho}^I  = D_{\rho\rho}+2 \epsilon_1 D_{\rho\omega} + \epsilon_1^2D_{\omega\omega}~,
\eeq
are second order in $\epsilon_{1,2}$ and will be neglected.

To introduce isospin-breaking most 
generally, we take for neutral and charged $\rho$ mesons independent $\rho\to\pi\pi$ 
couplings, $g_I\equiv g(\rho_I\to\pi^+\pi^-)$, $g_c\equiv
 g(\rho^+\to\pi^+\pi_3)$, and independent mass and width parameters entering $D^I_{\rho\rho}$ 
and $D^c_{\rho\rho}$. We neglect higher order effects in $g(\rho^+\to\pi^+\pi^0)$ caused by 
$\pi^0-\eta-\eta'$ mixing (\ref{mixing}),
\beq\label{grel}
g(\rho^+\to\pi^+\pi^0)=g(\rho^+\to\pi^+\pi_3)+\epsilon g(\rho^+\to\pi^+\eta)+\epsilon' g(\rho^+\to\pi^+\eta')~,
\eeq
because the two couplings multiplying $\epsilon$ and $\epsilon'$ violate G-parity and are
thus further suppressed; for instance~\cite{PDG}
\beq
\left|\frac{g({\rho^+\to\pi^+\eta})}{g({\rho^+\to\pi^+\pi_3})}\right|=\left[\left(1-\frac{m_\eta^2}{m_\rho^2}\right)
\frac{Br(\rho^+\to \pi^+\eta)}{Br(\rho^+\to\pi^+\pi_3)}\right]^{1/2}<0.055\quad (84\% {\rm CL})~.
\eeq

In the presence of isospin-breaking $\omega_I$ couples to two-pions with a coupling $g(\omega_I\to\pi\pi)$ 
of order $\epsilon_i g(\rho_I\pi\pi)$.
The decay $B\to(\pi\pi)^0X$ then proceeds either through $\rho_I$ or through $\omega_I$. 
Working to first order in isospin-breaking, these two contributions enter 
through a linear combination of the two propagators, both of order $\epsilon_i$,
\beq\label{tildeD}
\tilde D_{\rho\omega}(s)\equiv D_{\rho\omega}^I(s) +\frac{g(\omega_I\to\pi\pi)}{g(\rho_I\to\pi\pi)}
D_{\omega\omega}^I(s)~.
\eeq
Thus, one finds expressions for $B$ decay amplitudes into four pions, 
including terms which are first order in $\epsilon_i$:
\beqa\label{A+-}
A_{+-}(s_{12},s_{34})  & = & g_c^2 A(B^0\to\rho^+\rho^-)D_{\rho\rho}^c(s_{12}) D_{\rho\rho}^c(s_{34})~, \\
\label{A+0}
A_{+0}(s_{12},s_{34}) & = & g_cg_I \Big[ A(B^+\to\rho^+\rho_I)D_{\rho\rho}(s_{34})
\nonumber\\
& + &  A(B^+\to\rho^+\omega_I)\tilde D_{\rho\omega}(s_{34})\Big]
D_{\rho\rho}^c(s_{12})~,\\
\label{A00}
A_{00}(s_{12},s_{34})  & = & g_I^2 \Big [A(B^0\to \rho_I\rho_I)D_{\rho\rho}
(s_{12})D_{\rho\rho}(s_{34}) \nonumber\\
& + & \frac{1}{\sqrt{2}}A(B^0\to\rho_I\omega_I)
\Big (\tilde D_{\rho\omega}(s_{12})D_{\rho\rho}(s_{34}) + (s_{12}\leftrightarrow 
s_{34})\Big)\Big ]~.
\eeqa
An implicit angular dependence in \eqref{A+-}-\eqref{A00}, corresponding 
to given polarization states~\cite{BaBarbook}, is independent of $s_{12}$ and $s_{34}$.

In the isospin symmetry limit,  $D_{\rho\rho}^c = D_{\rho\rho},~\tilde 
D_{\rho\omega} =0$. Isospin-breaking is given by deviations from these 
equalities and, for the case of $\rho-\omega$ mixing,  is parametrized most generally 
by Eqs.~(\ref{A+-})-(\ref{A00}).  
Each term in a given row has a distinct dependence on $s_{12}$ and $s_{34}$,
characterized near the $\rho$ and $\omega$ poles by (\ref{Drho-omega}), (\ref{BW}) and 
(\ref{tildeD}). 
Taking $g_c/g_I = 1.005 \pm 0.010$~\cite{PDG},
the invariant mass distributions of the three processes permit in principle a determination of the 
three magnitudes, $|A(B^0\to\rho^+\rho^-)|, |A(B^+\to\rho^+\rho_I|$ and $|A(B^0\to \rho_I\rho_I)|$,
forming the isospin triangle, 
\beq\label{triangle-rho}
A(B^0\to \rho^+\rho^-) + \s A(B^0\to\rho_I\rho_I) - \s A(B^+\to\rho^+\rho_I) =0~.
\eeq
Once this triangle and its charge-conjugate are formed, one uses a phase relation for $A(B^\pm\to\rho^\pm\rho_I)$ analogous to (\ref{A+0phase})  
and the CP asymmetry in $B^0(t)\to\rho^+\rho^-$ to determine $\alpha$. This then provides
a way of extracting $\alpha$ free of effects from $\rho-\omega$ mixing. Electroweak 
penguin contributions are treated as in $B\to\pi\pi$, Eq. \eqref{DeltathetaEWP}.

The extraction of the pure isospin amplitudes $|A(B\to\rho\rho)|$ may be facilitated 
by using information from direct measurements of $A(B^+\to\rho^+\omega)$ and 
$A(B^0\to\rho^0\omega)$ entering the isospin-breaking terms in (\ref{A+-})-(\ref{A00}). 
Also, the isospin-breaking function $\tilde D_{\rho\omega}(s)$ is the same as the one 
fitted to the pion form factor~\cite{GO}. Denoting
\beq\label{info-pion-ff}
\tilde D_{\rho\omega}(s)=\tilde \Pi_{\rho\omega}(s) \frac{1}{[s-m_\rho^2+im_\rho\Gamma_\rho]} 
\frac{1}{[s-m_\omega^2+im_\omega\Gamma_\omega]}~,
\eeq
the fit yields $\tilde \Pi_{\rho\omega}(m_\omega^2)=-3500\pm 300~{\rm MeV}^2$, involving a 
possible small imaginary part compatible with zero. The slope at $s=m_\omega^2$, 
$\tilde \Pi_{\rho\omega}'(m_\omega^2)=0.03\pm0.04$, is consistent with zero. The exact $s$
dependence of $\tilde \Pi_{\rho\omega}$ is unimportant because its contribution is dominated by 
the narrow $\omega$ width. 

At the $\omega$ mass this gives 
\beq\label{Drho/omega}
\frac{|\tilde D_{\rho\omega}(m_\omega^2)|}{|D_{\rho\rho}(m_\omega^2)|} = 0.53 \pm 0.05~.
\eeq
Using the experimental values~\cite{rho+rho0,BaBaromega-rho},
\beq\label{BR-rho-omega}
{\cal B}(B^+\to\rho^+\rho^0) = (26.4^{+6.1}_{-6.4})\times 10^{-6}~,~~~~
{\cal B}(B^+\to \rho^+\omega) = (12.6^{+4.1}_{-3.8})\times 10^{-6}~,
\eeq 
and neglecting possible CP asymmetries in these processes, leads to
\beq\label{omega-rho-ratio}
\frac{|A(B^+\to \rho^+\omega)\tilde D_{\rho\omega}(m_\omega^2)|}
{|A(B^+\to\rho^+\rho^0)D_{\rho\rho}(m_\omega^2)|} =   0.36 \pm 0.08~.
\eeq
While this ratio becomes 0.02, typical for isospin-breaking, when weighed by the 
$\omega$ and $\rho$ widths, it has a large effect at the $\omega$ mass.

In order to demonstrate the effect of $\rho-\omega$ mixing 
on the $\pi^+\pi^-$ invariant mass distribution applying Eq.~(\ref{A+0}), one 
must use some information about the relative magnitudes and relative phases 
of the amplitudes for $B^{\pm}\to\rho^{\pm}\rho^0$ and $B^+\to\rho^{\pm}\omega$.
While the former amplitudes are pure ``tree" (we neglect very small EWP 
contributions), involving a single CKM phase ${\rm Arg}(V^*_{ub}V_{ud})=\gamma$, 
the latter involve also penguin contributions with weak phase ${\rm Arg}
(V^*_{cb}V_{cd}) = \pi$ in the $c$-convention~\cite{conv}, 
\beq
\s A(B^+\to \rho^+\rho^0) = {\rm t} + {\rm c}~,\quad 
\s A(B^+\to\rho^+\omega) = {\rm t} + {\rm c} +2{\rm p} + 2{\rm s}~.
\eeq
For our purpose, the terms ${\rm t}, {\rm c}, {\rm p}$ and ${\rm s}$ represent SU(3) 
amplitudes for longitudinally polarized vector mesons, similar to those defined 
in Sec. II for two pseudoscalars. The amplitude ${\rm s}$ is OZI-suppressed
and is expected to be negligible. (A similar amplitude in decays to a vector meson 
and a pseudoscalar meson dominates $B^+\to \phi\pi^+$ ~\cite{VP1}.)
The ratio $|{\rm p}|/|{\rm t} + {\rm c}|$ is small, about $0.1$, as can be inferred  
from the small branching ratio of $B^0\to\rho^0\rho^0$~\cite{BaBar}, or (by 
flavor SU(3)) from the measured longitudinal branching fraction of $B\to K^*\phi$~\cite{HFAG}.

\begin{figure}
\begin{center}
\epsfig{file=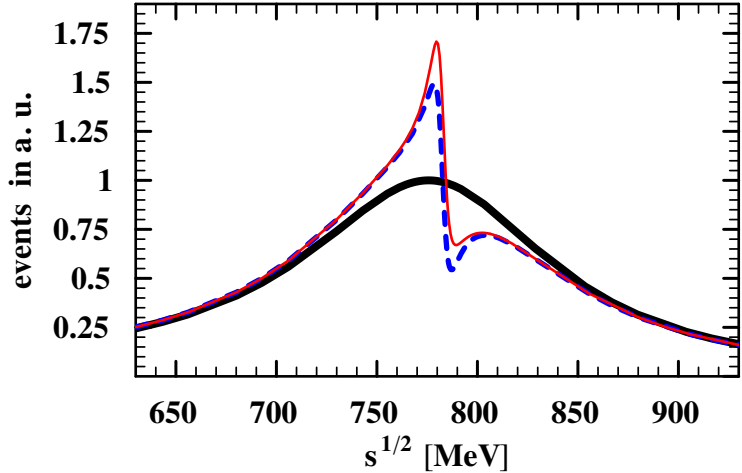}
\caption{Invariant mass distributions for $\pi^+\pi^-$ in $B^{\pm}\to\rho^{\pm}\pi^+\pi^-$ 
demonstrating $\rho-\omega$ mixing, using hadronic parameters as given in the text. 
Dashed (blue) line represents $B^+$ decays; solid (red) line represents $B^-$ decays; 
thick (black) line describes a case neglecting $\rho-\omega$ mixing.}
\label{Fig}
\end{center}
\end{figure}

For illustration, we use $2|{\rm s} + {\rm p}|/|{\rm t} + {\rm c}| = 0.2$, $\gamma = 57^\circ$, choosing
the strong phase difference between ${\rm s} + {\rm p}$ and ${\rm t} + {\rm c}$ to be zero,
so that the ratio ${\cal B}(B^+\to \rho^+\omega)/{\cal B}(B^+\to \rho^+\rho^0) = 0.82$ is smaller than one 
as implied by experiment, Eq.~(\ref{BR-rho-omega}). (Small transverse contributions are neglected.) 
These parameters determine relative magnitudes and relative phases between 
$A(B^{\pm}\to\rho^{\pm}\omega)$ and $A(B^{\pm}\to\rho^{\pm}\rho^0)$. The resulting effect on 
the $\pi^+\pi^-$ invariant mass distribution is shown in Fig.~1, separately for $B^+$ and $B^-$ decays. 
The vertical axis gives number of events in arbitrary units.
The prominent peak, followed by a dip, is characteristic of $\rho-\omega$ mixing~\cite{DKpipi}, 
and does not depend strongly on the choice of hadronic parameters as long as 
$|{\rm p}+{\rm s}|/|{\rm t}+{\rm c}| \ll 1$. Note that a small CP asymmetry is expected 
also in case that the strong phase difference 
between ${\rm s} + {\rm p}$ and ${\rm t} + {\rm c}$ 
vanishes, because of the two different 
shapes of the $\rho$ and $\omega$ resonances~\cite{ET,GOT,LGT}. 

In reality, limited statistics and particularly the current absence of a positive signal for 
$B^0\to\rho^0\rho^0$ would forbid carrying out the complete program leading to a 
construction of the pure isospin triangle (\ref{triangle-rho}). However, using the given 
invariant mass dependence of the isospin-breaking terms in (\ref{A+-})-(\ref{A00}), one 
may constrain these terms and eliminate them in certain cases. As noted above, a first place 
to look for these terms would be the $\pi^+\pi^-$ invariant mass distribution in 
$B^+\to\rho^+\pi^+\pi^-$ near the $\omega$ mass, 
which should be fitted to a sum of the $\rho^+\rho^0$ and $\rho^+\omega$ terms in (\ref{A+0})
as plotted in Fig.~1. 

In the relativistic Breit-Wigner form (\ref{BW}) we assumed no $s$ dependence in the 
width $\Gamma_V$. This assumption has only a very slight effect on the isospin-breaking
terms which are dominant at the narrow 
$\omega$ peak. Non-resonant contributions could also affect the invariant
mass dependence. One hopes to minimize these contributions by fitting the pion angular 
distributions to those describing longitudinally polarized vector meson states. 
Potential interference between the $\rho$ resonance and the wide $\rho'(1450)$ resonance 
must also be taken care of.

\section{Meson mixing and other isospin-breaking effects in $B\to \rho \pi$}

The isospin method for extracting $\alpha$ in $B\to\rho\pi$ is based on 
a time-dependent Dalitz plot analysis of $B^0\to\pi^+\pi^-\pi^0$~~\cite{QS}, using 
information provided by isospin symmetry~\cite{rhopi}. We will explain now the essence
of the method.  Consider the contributions of 
$\rho^+, \rho^-$ and $\rho^0$ to the amplitude of $B^0\to\pi^+\pi^-\pi^0$,
\beq\label{Drhorho}
A_{+-0}\equiv A(B^0\to\pi^+\pi^-\pi^0)=
A_+D_{\rho\rho}(s_+)\cos\theta_+ + A_-D_{\rho\rho}(s_-)\cos\theta_-+ A_0D_{\rho\rho}(s_0)\cos\theta_0~,
\eeq
where subscripts of the amplitudes $A_i$, the invariant masses $s_i$, and the helicity angles $\theta_i$, denote the charge of the $\rho$, 
\beq
A_+ \equiv A(B^0\to\rho^+\pi^-)~,~~~~A_- \equiv A(B^0\to\rho^-\pi^+)~,~~~
A_0\equiv A(B^0\to\rho^0\pi^0)~.
\eeq 
The function $D_{\rho\rho}$ is given near the $\rho$ pole by a Breit-Wigner 
form (\ref{BW}). Corresponding amplitudes for $\bar B^0$ are denoted by $\bar A$.
The time-dependent decay rate for an initially $B^0$ is given by~\cite{MG} 
\beqa\label{Gammat}
\Gamma(B^0\to\pi^+\pi^-\pi^0(t)) \propto (|A_{+-0}|^2 + |\bar A_{-+0}|^2) & + &  
(|A_{+-0}|^2 - |\bar A_{-+0}|^2)\cos(\Delta mt)\nonumber \\ 
& - & 2{\rm Im}\left ( e^{-2i\beta} \bar A_{-+0}A^*_{+-0}\right )\sin(\Delta mt)~.
\eeqa

The interference of the three $\rho$ resonances in the time-dependent and 
invariant mass-dependent decay rate permits a determination of the magnitudes of the 
three amplitudes $A_+,~A_-,~A_0$ and their charge conjugates, as well as the relative 
phases between these six amplitudes. This amounts to eleven independent observables,
obtained from a set of twenty seven mutually dependent measurables in 
(\ref{Gammat})~\cite{QS}.
It is convenient to rescale the $A_i$ and $\bar A_i$ amplitudes by phases $e^{i\beta}$  
and $e^{-i\beta}$, respectively, defining ${\cal A} _i\equiv \exp(i\beta)A_i,~\bar {\cal A}_i 
\equiv \exp(-i\beta)\bar A_i$, such that the coefficient of the $\sin(\Delta mt)$ term in  (\ref{Gammat})  
measures directly the relative phases between ${\cal A}_i$ and $\bar {\cal A}_i$.

Working in the $t$-convention~\cite{conv}, each amplitude $A_i$ consists of a 
``tree" contribution proportional to $V^*_{ub}V_{ud}$
and a ``penguin" term involving $V^*_{tb}V_{td}$.  
Unitarity of the CKM matrix is used to absorb matrix elements of penguin operators 
proportional to $V^*_{ub}V_{ud}$ in the tree amplitude,
\beq\label{decomp}
{\cal A}_{\pm,0}=e^{-i\alpha} T_{\pm,0}+P_{\pm,0}~,\qquad
\bar{\cal A}_{\pm,0}=e^{+i\alpha} T_{\pm,0}+P_{\pm,0}~,
\eeq
where $T_{\pm,0}$ and $P_{\pm,0}$ include strong phases.
The measured left-hand sides provide eleven equations for twelve unknown parameters, 
consisting of the weak phase $\alpha$, six magnitudes of tree and penguin amplitudes, 
and five relative strong phases between these amplitudes. An additional complex relation, 
leading to a total of thirteen equations for the twelve parameters, is provided by isospin 
symmetry. 
Neglecting electroweak penguin contributions and isopin breaking effects, the pure 
$\Delta I=1/2$ penguin terms vanish in the $I=2$ amplitude, implying
\beq\label{PenguinRel}
P_-+P_++2P_0=0~.
\eeq
The over-constrained set of equations \eqref{decomp}, \eqref{PenguinRel}  allows for 
the extraction of $\alpha$. The sensitivity to $\alpha$ can be seen explicitly by 
considering the $\Delta I=3/2$, $I=2$ amplitude ${\cal A}_2$,
\beq\label{T}
{\cal A}_2 \equiv {\cal A}_+ + {\cal A}_- + 2{\cal A}_0  = Te^{-i\alpha}~,\quad
\bar {\cal A}_2 \equiv \bar{\cal A}_+ + \bar{\cal A}_- + 2\bar{\cal A}_0 = Te^{i\alpha}~,
\eeq
with
$T\equiv T_+ +T_- +2T_0$.
The angle $\alpha$ is fixed by the relative phase between the two sums of amplitudes, 
which are determined up to an overall common phase by the time-dependent Dalitz plot.

The relations (\ref{PenguinRel}) and (\ref{T}) are violated both by EWP contributions and 
by isospin-breaking corrections caused, for instance, by $\pi^0-\eta-\eta'$ mixing and 
$\rho-\omega$ mixing
\cite{GM}. The former contributions may be included model-independently 
as in the case of $B\to\pi\pi$ and $B\to\rho\rho$, using the proportionality of the $\Delta I = 3/2$
current-current operator and the dominant $\Delta I =3/2$ EWP operator in the effective 
weak Hamiltonian~\cite{EWP,VPEWPCharles}. 
The relation (\ref{PenguinRel}) is modified to
\beq\label{PenguinRelEWP}
P_-+P_++2P_0=P_{EW}~,
\eeq
where $P_{EW}$ can be obtained from 
\beq\label{PenguinRelEW}
{\cal A}_2 =  Te^{-i\alpha} + P_{EW}~,
~~~~~
\frac{P_{EW}}{T} = -\frac{3}{2}\left (\frac{c_9 + c_{10}}{c_1 + c_2}\right )\frac{|V_{tb}V_{td}|}
{|V_{ub}V_{ud}|} = +0.013 \frac{\sin(\beta + \alpha)}{\sin\beta}~.
\eeq
This implies the same shift  in $\alpha$ as in $B\to\pi\pi$, $\Delta\alpha_{\rm EWP} 
= (1.5 \pm 0.3)^\circ$, if only the information from the two measured sums of amplitudes 
${\cal A}_2$ and $\bar {\cal A}_2$ is used to extract $\alpha$. 
The shift in $\alpha$ obtained from fitting the entire set~\eqref{decomp},
\eqref{PenguinRelEWP} and \eqref{PenguinRelEW} may be slightly different,
because of the additional $\alpha$ dependence of this over-constrained system 
of equations.

Isospin-breaking in  tree amplitudes does not affect the extracted value 
of  $\alpha$ in $B\to \rho\pi$, which is based on Eqs.~\eqref{decomp},~\eqref{PenguinRelEWP},
and \eqref{PenguinRelEW}, because no isospin relation between tree amplitudes is needed. 
This is obvious when $\alpha$ is extracted from (\ref{T}) and is true in general.
Penguin amplitudes in $B\to\rho\pi$ are known to be small~\cite{BN,PVSU3,GZ}. Therefore, 
other sources of isospin-breaking are expected to lead to corrections in $\alpha$ smaller than $\Delta\alpha_{\rm EWP}$, 
which is related to the $I=2$ tree amplitude through \eqref{PenguinRelEW}. The evaluation
of corrections caused by $\pi^0-\eta-\eta'$ mixing, that we give next, supports this expectation. 

The mixing of $\pi^0, \eta$ and $\eta'$ introduces additional correction terms in \eqref{T},
\beq\label{Tepsilon}
{\cal A}_+ + {\cal A}_- + 2{\cal A}_0  =  T e^{-i\alpha} + P_{EW}+ \epsilon 
P_{\rho\eta} +  \epsilon' P_{\rho\eta'}~, 
\eeq 
where $P_{\rho\eta^{(')}}$ are penguin amplitudes in $B^0\to \rho^0\eta^{(')}$. 
We expect that the correction to the extracted value of $\alpha$ is smaller in $B\to\rho\pi$ 
than in $B\to\pi\pi$ because the penguin-to-tree ratio is smaller in the first case. In terms of SU(3) amplitudes defined in~\cite{VP1,PVSU3}, one has
\beqa
|T| & = & |t_P + t_V + c_P + c_V|~,\nonumber\\
P_{\rho\eta} & = & \frac{1}{\sqrt{6}}(-p_P-p_V-s_V)~,\nonumber\\
P_{\rho\eta'} &= & \frac{1}{2\sqrt{3}}(p_P+p_V+4s_V)~.
\eeqa
A global SU(3) fit to available data of charmless $B$ decays to a pseudoscalar and a vector 
meson~\cite{PVSU3} has shown that $t_P$ and $t_V$ add up constructively, while $c_P$ and 
$c_V$ are smaller. Also,  $|p_P/t_P| \sim |p_V/t_V| \sim 0.2$~\cite{BN,GZ} and $p_V \simeq - p_P$, while $s_V$ is smaller. 
(A best fit gives~\cite{PVSU3} $|p_V/p_P|=1.15\pm0.07$, $\arg(p_V/p_P)=(182\pm18)^\circ$ and 
$s_V/p_V=0.16^{+0.08}_{-0.06}$.) All this implies that the effect of the terms in (\ref{Tepsilon}) 
involving $\epsilon$ and $\epsilon'$ is very small.
Taking
\beq\label{bound-ratios}
|T| \ge |t_V|~,~~~|P_{\rho\eta}|\simeq \frac{1}{\sqrt{6}}|s_V| \le \frac{0.3}{\sqrt{6}}|p_V|~,~~
|P_{\rho\eta'}| \simeq \frac{2}{\sqrt{3}}|s_V| \le \frac{0.6}{\sqrt{3}}|p_V|~,~~~\frac{|p_V|}{|t_V|} = 0.2~,
\eeq
and using $\epsilon = 0.017 \pm 0.003, \epsilon' = 0.004 \pm 0.001$,
we find the following upper bound on the uncertainty in $\alpha$ caused by neglecting the $\epsilon^{(')}$ terms in 
\eqref{Tepsilon}:
\beq
|\Delta\alpha_{\pi-\eta-\eta'}| =  \frac{|\epsilon P_{\rho\eta} +  \epsilon' P_{\rho\eta'}|}{|T|} \le
0.024\epsilon + 0.069\epsilon' \le  0.1^\circ ~.
\eeq 
In case that the sum of $p_P$ and $p_V$ in $P_{\rho\eta^{(')}}$ does not cancel 
completely~\cite{PVSU3}, the bound could be 
a factor two larger. In any event, this uncertainty is much smaller than 
$\Delta\alpha_{\rm EWP}$, the shift caused by EWP amplitudes.

Finally, we discuss the effect of  $\rho-\omega$ mixing treating it as in Sec.~III. 
Neglecting isospin-breaking in 
$g(\rho\to\pi\pi)$, we conclude that the third term in (\ref{Drhorho}) must be replaced by
\beq\label{rho0pi0}
A(B^0\to \rho^0 \pi^0)D_{\rho\rho}(s_0)\to A(B^0\to \rho_I \pi^0)D_{\rho\rho}(s_0)
+ A(B^0\to\omega_I\pi^0) \tilde D_{\rho\omega}(s_0)~,
\eeq
while the angular dependence remains unchanged. That is, 
the effect of $\rho-\omega$ mixing may be included in the time-dependent Dalitz 
plot analysis by adding the second term in (\ref{rho0pi0}). The isospin-breaking function 
$\tilde D_{\rho\omega}(s_0)$, defined in (\ref{tildeD}) and given in (\ref{info-pion-ff}), has a 
double pole structure with a narrow peak at the $\omega$ mass.  As discussed in the 
previous section, 
$\tilde D_{\rho\omega}(s_0)$ is measured by studying the pion form factor, 
while $A(B^0\to\omega_I\pi^0)$ can be  obtained from $B\to 4\pi$.
The narrow peak at the $\omega$ mass distinguishes clearly this isospin-breaking correction
from other potential non-resonant or wide resonance contributions to 
$B\to\rho\pi$~\cite{GM,nonresonant}.

The size of the effect of $\rho-\omega$ mixing may be estimated by considering the 
two processes, $B^0\to\rho^0\pi^0$ and $B^0\to\omega\pi^0$, occurring in (\ref{rho0pi0}).
The charge-averaged branching ratio of $B\to\rho^0\pi^0$ measured by Belle is somewhat 
larger than an upper limit reported by BaBar, reporting also the currently strongest upper bound on 
$B^0\to\omega\pi^0$,
\beqa
{\cal B}(B^0\to\rho^0\pi^0) & = & \left \{ \begin{array}{c}  (5.1 \pm 1.6 \pm 0.9)\times 10^{-6}~,~~~~
{\rm Belle}~~\mbox{\cite{Berho0pi0}}~, \cr
< 2.9\times 10^{-6}~,~~~~~~
{\rm BaBar}~~\mbox{\cite{Barho0pi0}}~, \end{array} \right.\nonumber\\
{\cal B}(B^0\to\omega\pi^0) &  & ~~~~~~~~< 1.2\times 10^{-6}~,~~~~~~{\rm BaBar}~~\mbox{\cite{Baomegapi0}}~.
\eeqa
These values and (\ref{Drho/omega}) permit a relatively sizable contribution from the
 isospin-breaking term $A(B^0\to\omega_I\pi^0) \tilde D_{\rho\omega}(s_0)$ at the pole, $s_0=m_\omega^2$,
 if ${\cal B}(B^0\to\omega\pi^0)$ is not much below its current upper limit~\cite{PVSU3}.

\section{Other sources of isospin violation}

In this work we have focused primarily on isospin-breaking effects in $B\to\pi\pi, B\to\rho\rho$
and $B\to\rho\pi$, originating in the mixing of neutral isospin triplet states ($\pi^0, \rho^0$) with 
isospin singlet states ($\eta^{(')}, \omega$). 
We also iterated  the effects of higher order electroweak penguin operators. 
Two implicit assumptions were made in our analysis:
\begin{itemize}
\item Reduced matrix elements of operators in the effective Hamiltonian, between initial 
$B^0$ and $B^+$ states and final states involving $\pi_3$ and $\pi^+$, were assumed to 
obey exact SU(2) relations. 
\item $\Delta I=5/2$ corrections were assumed to vanish.
\end{itemize} 
Relaxing these assumptions in  $B\to \pi\pi$ and $B\to \rho\rho$ introduces
 isospin-violating corrections in $\alpha$ which may be 
 comparable to those discussed in sections II and III. 
 $\Delta I = 5/2$ operators in $B\to\pi\pi$~\cite{GM} or in $B\to\rho\rho$~\cite{CKMfitter}
 may be induced by an insertion of the $d - u$ mass difference $\Delta I=1$ operator or by 
 electromagnetic corrections. This would violate the closure of  the isospin triangles 
 for $B$ and $\bar B$ amplitudes. Any of these other isospin-breaking corrections in 
 $B\to\rho\pi$ is expected to be negligible, however, because in these processes isospin 
 breaking can only 
 affect the relation \eqref{PenguinRelEWP} among suppressed penguin amplitudes. 
 
To see how these other effects enter a specific calculation, let us use the result of a Soft Collinear 
Effective Theory approach to factorization in $B\to M_1M_2$, where $M_{1,2}$ are 
pseudoscalars or vector mesons.
The result, at leading order in $\Lambda/m_B$,  is~\cite{BPRS}:
\beq
\begin{split}\label{SCETFact}
A=\frac{G_F m_B^2}{\sqrt{2}}\Big\{&f_{M_1}\int_0^1 du dz T_{1J}(u,z) 
\zeta_J^{BM_2}(z) \phi^{M_1}(u)\\
&+f_{M_1}\zeta^{BM_2}\int_0^1 T_{1\zeta}(u)\phi^{M_1}(u)\Big\}+
\Big\{1\leftrightarrow 2\Big\}+
\lambda_c^{(f)} A_{c\bar c}^{M_1,M_2}~.
\end{split}
\eeq
Here $T_{iJ}(u)$ and $T_{i\zeta}(u)$ are hard kernels which may be expanded in $\alpha_S(m_B)$, 
while $\zeta^{BM}$, $\zeta_J^{BM}(u)$ and the light cone meson wave function $\phi^M(u)$ are nonperturbative parameters. 
The amplitude $A_{c\bar c}^{M_1,M_2}$ 
denotes a possible long distance charming penguin contribution. 

We have quantified isospin-breaking caused by final states which do not 
coincide with isospin eigenstates. The remaining isospin violation is encoded in 
$\zeta^{BM}$, $\zeta_J^{BM}(z)$, $f_{M}\phi^M(u)$, and  $A_{c\bar c}^{M_1,M_2}$, 
where $M_{1,2}$ are now isospin eigenstates. Generically, the corrections are expected 
to be of order $(m_u-m_d)/\Lambda_{QCD}\sim \alpha_0\sim O(1\%)$, namely
of the same magnitude as the corrections caused by $\pi-\eta-\eta'$ mixing, Eq.~\eqref{bound'2},
and by EWP contributions, Eq.~\eqref{DeltathetaEWP}.
Isospin-breaking in the hard kernels $T_{iJ}(u), T_{i\zeta}(u)$ occurs through additional
$1/m_B$ power-suppressed operators and may be safely neglected. 
At this order, no isospin violation is caused by final state rescattering in the first two terms
in the amplitude \eqref{SCETFact}, because these terms  factorize to all orders in 
$\alpha_S$ and to first order in $\Lambda/m_B$. 

\section{Conclusions}

The extraction of the weak phase $\alpha \equiv \phi_2$ by application of isospin symmetry 
to $B\to\pi\pi$, $B\to\rho\pi$ and $B\to\rho\rho$ is modified through $\pi^0-\eta-\eta'$ 
mixing and $\rho-\omega$ mixing. We have studied these effects in a model-independent 
manner, discussing also other effects of isospin-breaking in these processes.
Our main results are the following:
\begin{itemize}
\item Isospin-breaking corrections in $\alpha$ related to $\pi^0-\eta-\eta'$ mixing were bounded 
using flavor SU(3), and were found to be smaller than $1.4^\circ$ in $B\to\pi\pi$ and much smaller 
in $B\to \rho\pi$. 
\item The effects of $\rho-\omega$ mixing in $B\to\rho\rho$ and $B\to\rho\pi$ were
studied as a function of the two-pion invariant mass in terms of a quantity measured 
in the pion form factor. Given the invariant mass dependence characterizing $\rho-\omega$ mixing,
which involves a peak at $s=m^2_\omega$, we propose a way for measuring and 
constraining these effects
experimentally.  Eventually, with sufficient statistics, this procedure may eliminate  
the mixing effect altogether.
\item In $B\to\rho\pi$, any kind of isospin-breaking in tree amplitudes does not affect the 
measurement of $\alpha$ through a time-dependent Dalitz plot analysis. 
Since penguin amplitudes are suppressed, the resulting uncertainty in $\alpha$ from 
isospin violation is expected to be smaller than one degree (excluding contributions
from EWP operators).
\item The proportionality of a $\Delta I =3/2$ current-current operator and a corresponding 
dominant electroweak penguin operator in the effective Hamiltonian implies a shift, 
$\Delta\alpha_{\rm EWP}=(1.5 \pm 0.3)^\circ$, common to $B\to\pi\pi$ and $B\to\rho\rho$.
The same shift would apply also to $B\to\rho\pi$ if only the sums of amplitudes (\ref{T}) were 
used. In a completely general extraction of $\alpha$ from the time-dependent Dalitz plot fit, 
the shift may be slightly different but can be obtained model-independently.
\end{itemize}
A brief summary of our conclusions is therefore: (1) Isospin-breaking introduces a much
smaller uncertainty in the value of $\alpha$ extracted from $B\to\pi\pi$ than thought before, 
of order $1^\circ$. 
(2) Effects of $\rho-\omega$ mixing in $B\to\rho\rho$ can be studied by fits to invariant mass 
distributions. (3) The largest shift in $\alpha$ in $B\to\rho\pi$, caused by electroweak 
penguin amplitudes, can be included model-independently, and is about $1^\circ$  as 
in $B\to\pi\pi$ and $B\to\rho\rho$.

\bigskip\medskip\noindent
{\bf ACKNOWLEDGMENTS}

\bigskip
We wish to thank Damir Be\' cirevi\' c, Frederic Blanc, Marko Bra\v cko, Svjetlana Fajfer, 
Andrei Gristan, Yuval Grossman,  Andreas Hoecker, Dan Pirjol, Jonathan Rosner, Ira 
Rothstein, Jim Smith, Denis Suprun and Alex Williamson for helpful discussions.
This work is partially supported by the Israel Science Foundation 
founded by the Israel Academy of Science and Humanities, Grant No. 1052/04,
and by the German--Israeli Foundation for Scientific Research and Development, 
Grant No. I-781-55.14/2003. The work of J. Z. is supported in part by the 
Department of Energy under Grants DOE-ER-40682-143 and DEAC02-6CH03000.

\appendix
\section{Experimental constraints on $\epsilon_{1,2}$}\label{appA}
Let us comment briefly on the values of the isospin-breaking parameters, $\epsilon_{1,2}$, 
which are constrained by fitting $\tilde D_{\rho\omega}(s)$ in \eqref{info-pion-ff} 
to the pion form factor~\cite{GO}. Requiring that $D_{\rho\omega}(s)$ does not have poles at 
$m_\rho^2-im_\rho\Gamma_\rho$ and $m_\omega^2- im_\omega\Gamma_\omega$ implies 
\beq
\epsilon_1=\frac{\Pi_{\rho\omega}(m_\omega^2-im_\omega\Gamma_\omega)}
{m_\omega^2-m_\rho^2 + i(m_\rho\Gamma_\rho - m_\omega\Gamma_\omega)}~, \qquad 
\epsilon_2=\frac{\Pi_{\rho\omega}(m_\rho^2-im_\rho\Gamma_\rho)}
{m_\omega^2-m_\rho^2 + i(m_\rho\Gamma_\rho - m_\omega\Gamma_\omega)}~.
\eeq
Using the relation
\beq\label{tilPi}
\tilde\Pi_{\rho\omega}(s)=\Pi_{\rho\omega}(s) +\frac{g({\omega_I\to\pi\pi})}
{g({\rho_I\to\pi\pi})}(s-m_\rho^2+ im_\rho\Gamma_\rho)~,
\eeq
one may express $\epsilon_i$ in terms of the measurable function $\tilde \Pi_{\rho\omega}(s)$:
\beqa
\epsilon_1 & = & \frac{\tilde\Pi_{\rho\omega}(m_\omega^2-im_\omega\Gamma_\omega)}
{m_\omega^2-m_\rho^2 + i(m_\rho\Gamma_\rho - m_\omega\Gamma_\omega)} -\frac{g({\omega_I\to\pi\pi})}{g({\rho_I\to\pi\pi})}~,\nonumber\\
\epsilon_2 & = & \frac{\tilde\Pi_{\rho\omega}(m_\rho^2- im_\rho\Gamma_\rho)}
{m_\omega^2-m_\rho^2 +  i(m_\rho\Gamma_\rho - m_\omega\Gamma_\omega)}~.
\eeqa 
The first term in $\epsilon_1$ is constrained experimentally~\cite{GO} (see also 
discussion below (\ref{info-pion-ff})),
\beq
\frac{\tilde\Pi_{\rho\omega}(m_\omega^2-im_\omega\Gamma_\omega)}
{m_\omega^2-m_\rho^2 + i(m_\rho\Gamma_\rho - m_\omega\Gamma_\omega)} 
=(-0.003\pm0.002)+ i (0.032\pm0.003)~.
\eeq
Barring a possible weak dependence of $\tilde \Pi_{\rho\omega}$ on $s$, this term is 
equal to $\epsilon_2$. The term $g({\omega\to\pi\pi})/g({\rho\to\pi\pi})$ in $\epsilon_1$  is 
poorly constrained experimentally, but is expected to be of the same order. The smallness 
of these parameters justifies neglecting terms of order $\epsilon^2_{1,2}$. Note that 
the method presented in Sec. III for studying isospin-breaking in $B\to\rho\rho$ depends
on the function $\tilde\Pi_{\rho\omega}(s)$ and not separately on its two components given
in (\ref{tilPi}).

\def \ajp#1#2#3{Am.\ J. Phys.\ {\bf#1}, #2 (#3)}
\def \apny#1#2#3{Ann.\ Phys.\ (N.Y.) {\bf#1}, #2 (#3)}
\def \app#1#2#3{Acta Phys.\ Polonica {\bf#1}, #2 (#3)}
\def \arnps#1#2#3{Ann.\ Rev.\ Nucl.\ Part.\ Sci.\ {\bf#1}, #2 (#3)}
\def \art{and references therein}
\def \cmts#1#2#3{Comments on Nucl.\ Part.\ Phys.\ {\bf#1}, #2 (#3)}
\def \cn{Collaboration}
\def \cp89{{\it CP Violation,} edited by C. Jarlskog (World Scientific,Singapore, 1989)}
\def \econf#1#2#3{Electronic Conference Proceedings {\bf#1}, #2 (#3)}
\def \efi{Enrico Fermi Institute Report No.}
\def \epjc#1#2#3{Eur.\ Phys.\ J.\ C {\bf#1}, #2 (#3) }
\def \ib{{\it ibid.}~}\def \ibj#1#2#3{~{\bf#1}, #2 (#3)}
\def \ijmpa#1#2#3{Int.\ J.\ Mod.\ Phys.\ A {\bf#1}, #2 (#3)}
\def \ite{{\it et al.}}
\def \jhep#1#2#3{JHEP {\bf#1}, #2 (#3)}
\def \jpb#1#2#3{J.\ Phys.\ B {\bf#1}, #2 (#3)}
\def \jpg#1#2#3{J.\ Phys.\ G {\bf#1}, #2 (#3)}
\def \kdvs#1#2#3{{Kong.\ Danske Vid.\ Selsk., Matt-fys.\ Medd.} {\bf #1}, No.\#2 (#3)}
\def \mpla#1#2#3{Mod.\ Phys.\ Lett.\ A {\bf#1}, #2 (#3)}
\def \nat#1#2#3{Nature {\bf#1}, #2 (#3)}
\def \nc#1#2#3{Nuovo Cim.\ {\bf#1}, #2 (#3)}
\def \nima#1#2#3{Nucl.\ Instr.\ Meth.\ A {\bf#1}, #2 (#3)}
\def \npa#1#2#3{Nucl.\ Phys.\ A~{\bf#1}, #2 (#3)}
\def \npb#1#2#3{Nucl.\ Phys.\ B~{\bf#1}, #2 (#3)}
\def \npps#1#2#3{Nucl.\ Phys.\ Proc.\ Suppl.\ {\bf#1}, #2 (#3)}
\def \PDG{Particle Data Group, S. Eidelman \ite, \plb{592}{1}{2004}}
\def \pisma#1#2#3#4{Pis'ma Zh.\ Eksp.\ Teor.\ Fiz.\ {\bf#1}, #2 (#3) [JETPLett.\ {\bf#1}, #4 (#3)]}
\def \pl#1#2#3{Phys.\ Lett.\ {\bf#1}, #2 (#3)}
\def \pla#1#2#3{Phys.\ Lett.\ A {\bf#1}, #2 (#3)}
\def \plb#1#2#3{Phys.\ Lett.\ B {\bf#1}, #2 (#3)} 
\def \ppnp#1#2#3{Prog.\ Part.\ Nucl.\ Phys.\ {\bf#1}, #2 (#3)}
\def \prd#1#2#3{Phys.\ Rev.\ D\ {\bf#1}, #2 (#3)}
\def \prl#1#2#3{Phys.\ Rev.\ Lett.\ {\bf#1}, #2 (#3)} 
\def \prp#1#2#3{Phys.\ Rep.\ {\bf#1}, #2 (#3)}
\def \ptp#1#2#3{Prog.\ Theor.\ Phys.\ {\bf#1}, #2 (#3)}
\def \rmp#1#2#3{Rev.\ Mod.\ Phys.\ {\bf#1}, #2 (#3)}
\def \rp#1{~~~~~\ldots\ldots{\rm rp~}{#1}~~~~~}
\def \yaf#1#2#3#4{Yad.\ Fiz.\ {\bf#1}, #2 (#3) [Sov.\ J.\ Nucl.\ Phys.\ {\bf #1}, #4 (#3)]}
\def \zhetf#1#2#3#4#5#6{Zh.\ Eksp.\ Teor.\ Fiz.\ {\bf #1}, #2 (#3) [Sov.\Phys.\ - JETP {\bf #4}, #5 (#6)]}
\def \zp#1#2#3{Zeit.\ Phys.\  {\bf#1}, #2  (#3)}
\def \zpc#1#2#3{Zeit.\ Phys.\ C {\bf#1}, #2 (#3)}
\def \zpd#1#2#3{Zeit.\ Phys.\ D {\bf#1}, #2 (#3)}



\begin{thebibliography}{99}
\bibitem{identical} Field operators creating a state consisting of two identical
particles can do this in two different ways implying a factor of 2
in the amplitude.  However, when integrating over the final states, one has
to avoid double counting of configurations with identical momenta, so the
rate is cut by a factor of 2 and the net result is a factor of $\s$ in the amplitude.
\bibitem{Delta-phi} The correction to the angle $\phi$ in (5) is given by 
$\Delta \phi= (e_0/\sin\phi)(\sqrt{2}|A_{+0}|/|A_{+-}| -\cos\phi)$. A similar 
expression describes the correction to the angle $\bar\phi$.
\end{thebibliography}
\end{document}